# *Non*-Concept Software Subsystems: Tangible and Intangible

**Iaakov Exman**

Software Engineering Department
The Jerusalem College of Engineering – JCE – Azrieli
POB 3566, Jerusalem, 91035, Israel
iaakov@jce.ac.il

**Categories and subject Descriptors**

D.2.2  [Software Engineering]: Design Tools and Techniques {1998 ACM Classification}

Software and its engineering → Software creation and management → Designing software {CCS 2012}

**Abstract**:    Concepts modified by a *Non-* prefix apparently denote a negation, an opposite of the concept without this prefix. But, generally the situation is rather subtle: *non-* implies only partial negation and the *concept* suggests preserved identity with some reduced quality or absent attribute. In this work tangible and intangible software subsystems based upon *Non-* concepts are defined and pluggable ontologies are proposed for their representation. Pluggable ontologies are a kind of nano-ontologies, which by their minimal size facilitate fast composition of new software subsystems. These ontologies are made pluggable by Design Sockets, a novel kind of class. These are abstract connectors for removed/added parts, functionalities or identities, and for subdued qualities. Design Sockets are the basis of a Design Pattern for dynamically modifiable software systems. Pragmatic implications of *Non-* concepts include manageable design of product lines with multiple models. *Non-* concepts are also relevant to the controversy whether composition is/isn't identity. The resolution is not sharp. Identity is entangled with composition, and is preserved to a certain extent, until further removal causes identity break-down.

## 1   INTRODUCTION

The top-level concept of an ontology is called *thing*. All software subsystems are things. This is a similar idea to the Java programing language root-class called *Object*, which is a superclass of all other classes in the language. A next level pair of concepts of an ontology is *tangible* and *intangible*. An informal way of defining a *tangible thing* is to say that it is physical, e.g. a car. An *intangible thing* is not physical, e.g. an algorithm or a nostalgia feeling. *Non-*concepts, soon to be motivated, can also be of two types: tangible and intangible.

### 1.1   Software Systems are expressed in Natural Language Concepts

Our basic assumption is that software systems in their highest abstraction design levels are expressed in terms of natural language concepts – and not programming language constructs. The relevant structures in these abstraction levels are application ontologies [15], [19] which are specialized for specific applications and much smaller than domain ontologies. From application ontologies one is able to generate [14] the next downward abstraction level, viz. UML class diagrams, and so on down to the code executable in a machine, real or virtual.

In order to redesign a software system in the highest abstraction levels, efficient mechanisms are needed for modifying subsystems. Efficiency requires usage of smaller nano-ontologies [12] with a few ontology classes.

Enter *Non*-concepts.

*Intangible non-concepts* – a concise expression for "non-concepts that refer to intangible things" – are used in colloquial natural language and triggered the idea of non-concept subsystems. Intangible non-concepts refer to notions with some less intense or almost absent quality. For instance, one possible meaning of *non*-trivial (often written without a hyphen) is a *thing of some importance*. This usage is different from a logical not: it is neither saying that the referred thing is not important, nor that it is decidedly important. Although dictionary-wise "non-" is a prefix indicating negation, we shall use it in this work as a noun, with the specific meaning of a kind of concept.

*Tangible non-concepts* refer to objects with some part removed/added or some functionality lost. For instance, a *non*-clock could be an electronic clock with a battery removed, so that it cannot perform its essential function of showing the current time.

*Non*-concepts are an easy way to make slight modifications in a well-known system. Instead of reproducing a whole (possibly large) application ontology for the system with small additions/changes/deletions here and there, it suffices to represent the modified system by a single concept (class) standing for the whole original system and the very few explicit changes that are performed. This kind of nano-ontology is very concise. We build these nano-ontologies with a *pluggable* structure – first introduced in ref. [11] – based upon abstract connectors enabling varying changes on the same system as desired, say to build a product line.

This paper explores the space of possible *Non-* concepts touching related conceptual issues. Concerning tangible non-concepts, when an object is stripped of some of its parts or loses functionality it may reach a point where it is not anymore recognizable as such a kind of object: besides its utility, it loses its identity. But, there are intriguing situations in which a loss of functionality or parts does not lead to loss of identity. A concept assigned to an object in such situation is a "*Non-*" concept. It is both *Non-* as it has lost some of its characteristics, and still a *concept* as it is easily recognizable as such by our senses (vision, hearing, etc.). Concerning intangible non-concepts, the issue of identification or recognition is of less importance. Their usage in colloquial language is to qualify – i.e. to assert the modified quality of – the referred concept.

## 1.2  Paper Organization

The remaining of this paper is organized as follows. Section 1 is concluded with a Related Work sub-section. Section 2 introduces tangible non-concepts, using an example. Section 3 defines pluggable ontologies for tangible non-concepts. Section 4 examines spaces of non-concepts. Section 5 makes a transition between tangible and intangible non-concepts with some examples from art. Section 6 deals with intangible non-concepts. Section 7 summarizes non-concepts with a dynamically modifiable Design Pattern. The paper is concluded by a discussion in Section 8.

## 1.3  Related Work

A gentle introduction to formal ontologies – used in this work to represent Non- concepts – can be found in e.g. Guarino [19]; see also Bacon et al. [3]. Nano-ontologies were introduced in ref. [12] and were applied within a location-based recommendation system by Exman and Nagar [13].

Modular ontologies – composed of sub-ontologies – have been proposed and extensively discussed. A few representative pointers include Rector et al. [28], Schlicht [30] and Hois et al. [20].

Physical objects – which are relevant to our tangible non-concepts – have been dealt with by Borgo, Guarino and Masolo [7]. Their discussions, among other topics, distinguish between matter and physical object, and ask whether a broken cube, which is nonetheless recognized as a cube, is indeed still a cube.

The negation challenge of a concept within Formal Concept Analysis (FCA) has been considered, for instance by Priss [26] who asks the meaning of a negated concept such as "not a piano" or a negated attribute such as "not green". A possible meaning for such a negation could be given by the complement with respect to a super-concept, which is not always available or reasonable. In particular with respect to concept lattices, it may be the case that negations of formal concepts are not formal concepts themselves. Baader et al. [2] deal with partial contexts and negation of concepts when using FCA to complete Description Logic (DL) knowledge bases. Ferre [16] differentiates between Negation, Opposition and Possibility. For example, negation is "old/not old", while opposition is "old/young". Negations and changes are also dealt with by Flouris et al. [17].

Other meanings of non-concept have been found in the literature. For example, Reiterer et al. [29] refer to non-concepts that are the complete opposite of certain concepts. The nature of this opposite concept is not specified.

Non- concepts do not imply malfunction, defective or broken objects, or incomplete and/or inconsistent ontologies. Design problems leading to incomplete and/or inconsistent ontologies have been dealt with in the literature, e.g. Baumeister et al. [6] and Hwang [21].

Identity has been a widely discussed issue e.g. Kripke [22], and often relates identity to composition. There are two roughly opposing positions with this respect. In one position identity *is* composition of parts. Some representative examples are e.g. Lewis [23], Merricks [25] and Liao [24]. Lewis states that the opposite of identity is not non-identity, but distinctness in the sense of overlap, things with parts in common (see [23] page 33). This is a suitable starting point for this work.

In the other camp a set of variations on the composition *is not* identity. See e.g. Baker [4] and Elder [9]. Inquiring deeper one finds that both camps have more in common than acknowledged.

General references on mereology – the study of parts of a system – are e.g. Simons [31] and Varzi [32].

Systems' functionality or behavior has been less under the focus of conceptual approaches.

## 2  TANGIBLE *NON*-CONCEPTS

The first part of this paper refers to tangible non-concepts that will be used to represent tangible systems and their corresponding software subsystems.

### 2.1  The Non-Clock Example

The author of this paper has a non-clock hanging on a wall in the kitchen. It is seen in Fig. 1. It is used to illustrate the idea of non-clock for curious guests. It cannot be used to measure time.

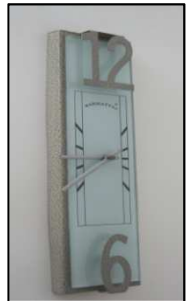

Figure 1: Photo of a Non-Clock – A non-clock as a concrete instance of its *Non-* concept. It has a visible scale – the numbers 6, 12, and marks for other hours. It is synchronizable by a mechanism in its back. It lacks periodicity since its battery was removed. Its identity is clearly recognized, but it is not useful for measuring time.

In order to generate a meaningful non-concept for a tangible subsystem one first needs a clear conceptual characterization of the subsystem essence. One needs to know the essential functionalities without which it is not considered anymore an instance of such a type of subsystem.

For instance, we have characterized in [10] a clock as a device to measure time with three essential conceptual properties:

1. <u>Periodicity</u> - it has a periodic behavior, based upon a physical phenomenon;
2. <u>Adjustability</u> – it has a pre-defined scale of numbers, to which events are assigned;
3. <u>Synchronization</u> – it may send/receive messages, to synchronize with other clocks.

The non-clock in Fig. 1 has a scale as clearly seen: the numbers 6 and 12 and marks for other hours. It may be synchronized and is adjustable, since one can rotate the non-clock hands to any desired value in the scale, by a mechanism in its back. It does not have periodicity, since its battery has been removed. Thus, it cannot be used to measure time. It lacks both a component part and its correspondent functionality.

Nonetheless, one easily recognizes its identity. Ask any guest – what is hanging on the wall? – and one easily gets a "clock" reply. It takes some time to explain that it is a non-clock.

The object in Fig. 1 is not a ¬clock, where ¬ is the logical not sign. The referred object is not the complement of a clock in any chosen universe of objects. The very fact of its recognition implies that it is much closer to be a clock than whatever may be its complementary ¬clock. A non- concept is neither the (original) concept, nor the ¬concept (logical negation of the original concept).

## 2.2 Tangible Non- Concepts Defined

We define a Tangible Non- concept as follows.

> **Tangible Non- Concept:** Definition
>
> A Tangible *Non*-concept is assigned to a sub-system, when it is empirically verifiable that:
> a. Some of the sub-system's parts and/or functionalities are removed/added;
> b. The sub-system keeps the identity of the respective concept, i.e. the referred removal/additions do not affect the identification capability.

In the above definition there are four elementary undefined concepts:

a) *Identity* – there may be several object identities, but just a single unified identity in a given context;
b) *Part* – a discrete structural component of the sub-system that may be added or removed;
c) *Functionality* – a discrete behavior of the sub-system, associated with one or more of its parts;
d) *Non-* – a noun serving as a special kind of identity of a concept.

Tangible non-concepts do not refer to gradual change. They mean discrete removal/addition of parts or functionalities, leading to a distinct entity of a new kind. One can remove/add parts without affecting identification. In fact, there exist products explicitly designed to allow such removal/addition. Nevertheless in each object there exist essential parts that once removed prevent identification of the original object. Functionalities are quite similar to component parts. Removal/addition of functionality does not necessarily prevent identification.

The identification issue – with its philosophical connotation – is indeed relevant to the structural and functional meaning of software subsystems. The idea of *conceptual integrity* has been emphasized by Brooks [8] as the most important consideration in system design. Thus, it seems that one still recognizes a remaining system by means of its conceptual integrity, even after removal of certain parts.

## 3    PLUGGABLE NANO-ONTOLOGIES

Pluggable nano-ontologies (see [12]) are very small size ontologies with inherent preparations for dynamically modifiable composition, i.e. parts or functionalities may be plugged-in or out. The inherent plugging points are called *Design Sockets*, a special purpose kind of class, to be added to ontologies in order to represent *Non-*concepts.

Design Sockets solve the following problem: -How to fully represent a subsystem's *Non-* concept displaying the whole original subsystem as a single concept (class) and with just a few removable parts which have been actually removed/restored/added?

### 3.1 Design Sockets

*Design Socket* is an abstract generic connector for any of the above concepts: identity, part, functionality. It allows dealing in a neat way with removal and addition of identities, sub-system parts and functionalities. A Socket is itself a class. Each Socket has one or more "plugged-in" concepts (classes or properties), the respective pluggable parts or functionalities.

A plugged-in concept (class) has a cardinality restriction whose value is Boolean. A part or functionality is either plugged-in with cardinality value 1, or not plugged-in, with value 0. In fact, the pluggable ontology with plugged-in classes can be viewed as a kind of UML class diagram in which "plugged-in" is the name of an association between pairs of classes and the cardinality refers to the endpoint near each specific property.

### 3.2 iSockets

An *iSocket*, standing for identity socket, is a sub-class of Socket as seen in Fig. 2, specialized for *identity* removal/addition. The cardinality of the plugged-in property of an iSocket is omitted, as it is always 1. The inheritance arrow from an iSocket to the Class Socket is also omitted in the nano-ontologies of subsystems.

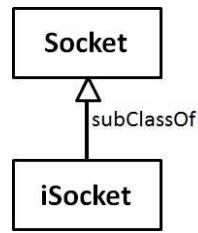

Figure 2: iSocket – a sub-class of Socket specialized for subsystem identity. *Non-* is a pluggable concept of iSockets.

*Non-* is only used as a concept that is pluggable in iSockets. There may be only a single *Non-* in the iSocket of an object. This is different from the logical not sign ¬ which can be added to each proposition, thus appear several times in the description of a single object.

### 3.3 Pluggable Ontology: Tangible Non-Concept Examples

We start the tangible examples with the non-clock of sub-section 2.1. Its battery was removed, thus it has no periodicity. Since the adjustability and synchronization were not modified, they are not represented. This explains why the pluggable ontology is so small: one needs to explicitly represent only the modifiable properties. The non-clock pluggable ontology is in Fig.3. All the arrows in this and subsequent pluggable ontologies – with rhombus (lozenge) arrow heads – represent *composition* (*isPartOf*) arrows. Inheritance arrows (subClassOf) are omitted for clarity.

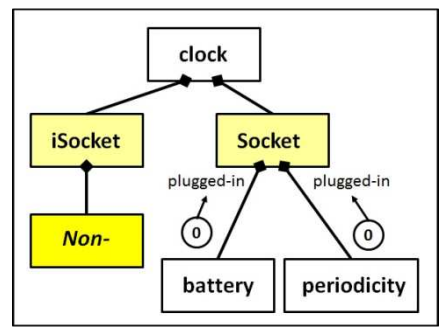

Figure 3: Non-clock pluggable ontology – It has one iSocket with a *Non-* value. It is a non-clock as its battery was removed: consequently it also lost its periodicity functionality. Cardinality values of both plugged-in properties are 0, as the respective part and functionality were removed.

Let us perform a thought ("gedanken") experiment. Suppose we add a new battery to our non-clock. We then synchronize and adjust the time shown to be the correct current time. So, now it is just a fine functioning clock. Next we put an internet video camera in front of the revived clock. The image of the moving clock is transmitted through the internet, and seen in another computer – in a different country.

The image of clock through the internet is now an Internet-Video non-clock. It has a scale and periodicity. Its identity is easily recognized as an instrument to measure time and can be used to do so.

But the video clock itself cannot be remotely synchronized through the video screen, unless we purposely add special software to this end. So, by the demand of the three properties, in sub-section 2.1 above, it is not a plain clock. Nonetheless it is a useful non-clock, as long as the actual clock which is the video image source works well. Its pluggable ontology is seen in Fig. 4. It differs from the previous ontology by a plugged-in addition – the internet-video – and two plugged-in removals: the synchronization functionality and the subsystem part responsible for it, called Synch-part.

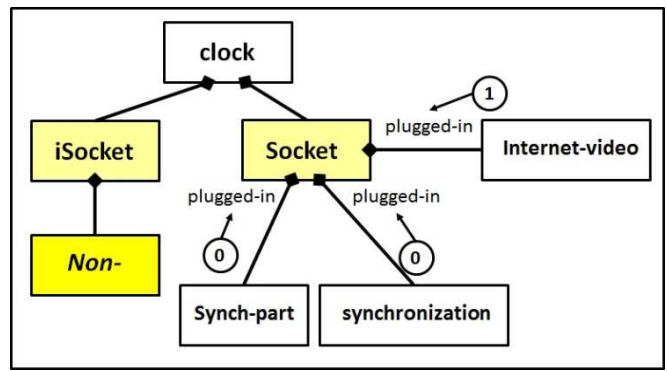

Figure 4: Internet-Video Non-clock pluggable ontology – It is a non-clock as we see its video through the internet: it lost its synchronization functionality. The cardinality values of the plugged-in synch-part and synchronization functionality are 0. On the other hand, an internet-video property has been added with cardinality 1. The periodicity property is omitted as it was restored to the clock and one does not leave it modifiable.

Next, we perform a second thought experiment. We keep the internet video camera, but again remove the clock's battery. The video image now is static. This new non-clock is not very useful. It certainly has a scale, but no periodicity and no synchronization ability.

## 4  THE NON-CONCEPTS SPACE

### 4.1 Non-Concepts by Design

There are products a priori designed to fit *Non-* concepts, those displaying:

   a) *lacking parts*;
   b) *downgraded components*;
   c) *lacking functionalities*.

For instance, most portable electronic devices are sold without batteries, which are added in the purchase act. Printers are sometime sold with downgraded toner containers (with less toner quantity) (see Fig. 5).

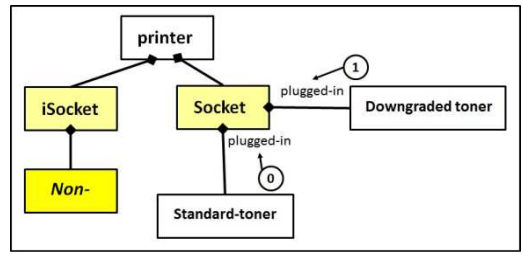

Figure 5: Non-printer with downgraded toner pluggable ontology – It is a non-printer since the standard toner was removed – thus it has plugged-in cardinality=0. It is sold with downgraded toner – its corresponding plugged-in cardinality=1.

## 4.2   Non-Concepts, Obsolescence and Their Cemeteries

A natural kind of non-concepts is obtained by aging or obsolescence. Some of these objects e.g. cars are thrown away in cemeteries, see Fig. 6. They have been the subject of literary works and a theatre play. A non-car's ontology for a car rescued from the cemetery to be a collector's item is shown in Fig. 7.

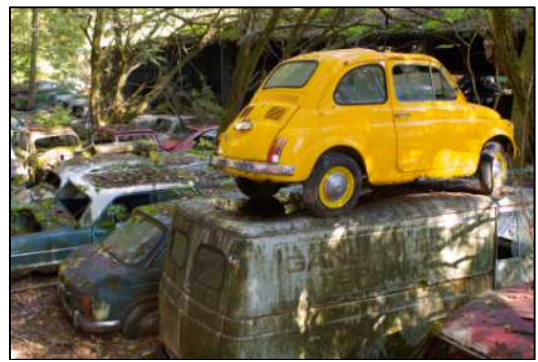

Figure 6: Yellow non-car in Car cemetery photo – All the cars in a cemetery are identifiable as non-cars. One cannot tell through the photo that the yellow one is for sure a non-car, but its overall condition leads us to think so.
Photo: Norbert Aepli, published under the "Creative Commons Attribution 3.0" license.

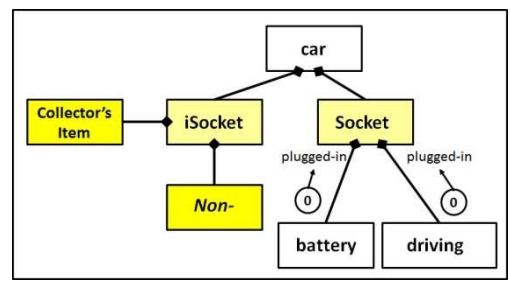

Figure 7: Collector's Non-car pluggable ontology – It is a non-car as the driving functionality was removed, say by removing the battery, both with plugged-in cardinality=0. It serves as a collector's item, as shown plugged-in in the iSocket.

The recent fast technological evolution may turn a perfectly working and useful object into a non-object thrown away to less fashionable cemeteries. There are uncountable instances. The typewriter, the slide rule calculator, the camera with chemical film, and the CRT computer screen, were all displaced by disrupting technologies. All their components may be in place and with flawless original functionality when discarded. Note that non-cars actually refer to single instances, while the non-typewriters' obsolescence refers at once to the whole class of such objects.

## 5   TRANSITION FROM TANGIBLE TO INTANGIBLE *NON*-CONCEPTS: ART OBJECTS

In this section we deal with art objects which are still tangible, but apparently have less the character of an artefact subsystem. Thus they are a sort of transition to purely intangible non-concepts.

### 5.1   Magritte's *Non*-Pipe

The well-known art work "Non-Pipe" by the Belgian surrealist Rene Magritte is seen in Fig. 8. It has inscribed in French "Ceci n'est pas une pipe", which means "This is not a pipe".

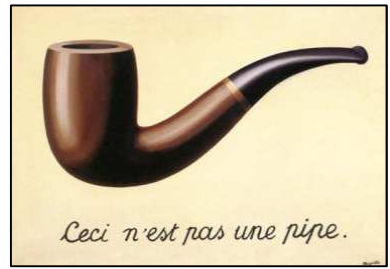

Figure 8: "Non-pipe" by Magritte – this is an image of the art work by the Belgian surrealist painter. In the work itself the object is called a non-pipe. One can only appreciate its message if one recognizes the identity of the object as a pipe. Thus, it is an actual non-pipe.

Recognizing the identity of the non-pipe as a pipe is essential to understand the witty message. It really is a non-pipe since its functionality is removed – one cannot smoke with a non-pipe – similarly to our non-clock photo discussed in sub-section 2.1.
The non-pipe ontology is shown in Fig. 9. It removes the pipe 3-dimensionality and adds 2-dimensionality.

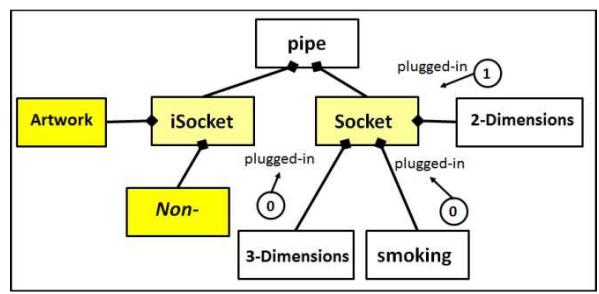

Figure 9: Artwork Non-pipe pluggable ontology – It is a non-pipe since the smoking functionality was removed, by removing the 3-dimensions – plugged-in cardinality=0 while adding 2-dimensions. It serves only as an artwork – as shown plugged-in in the iSocket.

In an age in which printers can print 3-dimensional objects, it is not too far-fetched to imagine Magritte's non-pipe image as an input to such a printer, while the output could be smokable.
Continuing with the idea of Internet-Video – used for a clock, in Fig. 4, sub-section 3.3 – one could perform the following thought experiment. A person is smoking a pipe in a certain location, and this scene can be viewed by means of internet-video in a remote location. This simplistic remote video of the pipe is indeed with 2-Dimensions and non-smokable at a distance. This modified system in this video experiment could be shown to be smoking despite being 2-Dimensional, and together with transmission of odours through the internet, one could even feel the smoke smell.

## 5.2  Magritte's *Non*-Apple

Magritte also created other non-objects, such as a non-apple with the analogous sentence "This is not an apple." seen in Fig. 10. It is a non-apple as it lacks many of the behaviours and functionalities of a natural apple, e.g. edibility.

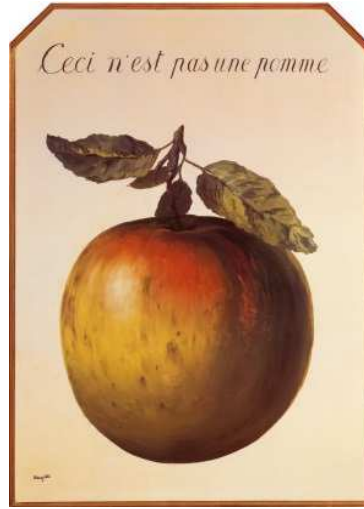

Figure 10: "Non-apple" by Magritte – this is an image of the art work by Magritte, similar in spirit to the non-pipe in Fig. 8. In the work itself the object is so-to-speak called a non-apple (in French). Thus, it is an actual non-apple.

The non-apple ontology, very similar to the non-pipe ontology, is shown in Fig. 11.

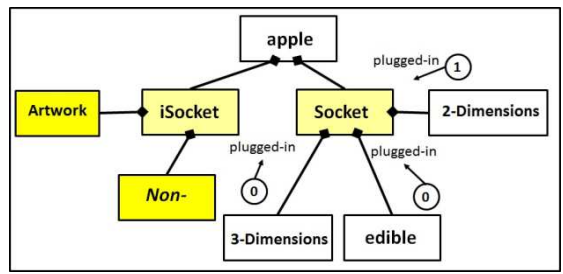

Figure 11: Artwork Non-apple pluggable ontology – It is a non-apple since e.g. the edibility functionality was removed. Compare with the non-pipe in Fig. 9.

The non-apple is decidedly not an artefact produced by humans. It is a non-concept of a subsystem only in a biological context. In this sense it is closer to an intangible non-concept. One could repeat the Internet-Video experiment with a real apple in a remote location. Then the apple would be edible, and could be actually eaten by someone like a "reality" television show.

Finally, a non-Apple that could not be conceived by Magritte is shown in Fig. 12. Magritte died in 1967 and the Apple Company was founded only in 1976.

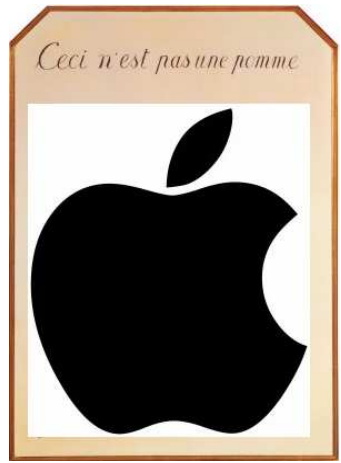

Figure 12: "Non-Magritte-non-Apple" – this is a non-Apple image of a non-Magritte "Logowork", paraphrasing the non-apple in Fig. 10. In the work itself the object is so-to-speak called a non-apple (in French). Thus, it is an actual non-Magritte.

Fig. 12 needs some explanation. In colloquial language one refers to a Picasso painting, as "a Picasso". Thus, Fig. 10 can be said to be an image of "a Magritte", but not Fig. 12, although one may recognize underlying Magritte characteristics. In this sense Fig. 12 is a Non-Magritte, besides still being a non-apple. As an aside, Fig. 12 rather than an artwork, it is a "Logowork".

The Non-Magritte-non-apple ontology, similar to the previous non-apple ontology, is shown in Fig. 13.

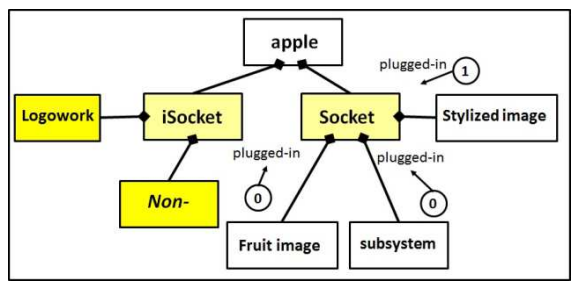

Figure 13: Non-Magritte-Non-apple pluggable ontology – It is a non-apple by the same reason of the non-Apple in Fig. 11. In addition, the fruit image was deleted, thus edibility is not an issue at all. A stylized image (a computer logo) was plugged-in, partially recovering a possible artefact subsystem. It is a Non-Magritte since one still recognizes the Magritte style through the French sentence and the same frame of Fig. 10.

### 5.3   Altman's *Non*-Person

Still another artwork example is the *non-* person idea, used by the film director Robert Altman in his Gosford Park [1] movie from 2001 to describe social classes in an aristocratic mansion in Britain. Each of the aristocrat characters has a personal servant.

The personal servants are not called by their proper names, but rather by the names of their respective owners, Mr. such and such. The servants, obviously recognized as persons, are deprived of the most elementary right of being called by their own private names. The simultaneous recognition of their identity and lack of identity is a very effective means to stress social strata. The respective pluggable ontology is shown in Fig. 14.

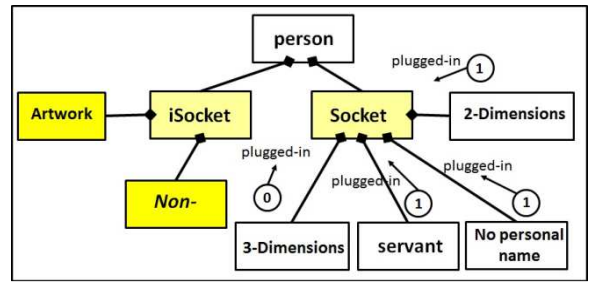

Figure 14: Artwork Non-person pluggable ontology – It is a non-person since the 3-dimension property was removed, while adding 2-dimensions, being a servant and without personal name. It serves as an artwork – as shown plugged-in in the iSocket.

## 6   INTANGIBLE *NON*-CONCEPTS

We define an Intangible Non- concept as follows:

> **Intangible Non- Concept:** Definition
>
> An Intangible *Non*-concept is assigned to a composed term, when:
> c. The composed term has a prefix *Non-* prepended to an intangible original term;
> d. The composed term both keeps a positive property common with the intangible original term and additionally displays a negative property.

In order to illustrate this definition we give in the next sub-section a few examples, with their respective pluggable ontologies.

### 6.1   Pluggable Ontology for Intangible Non-Concepts

Besides the *Socket* and *iSocket* classes defined above in section 3, we introduce here just two minimal additional classes needed to represent intangible Non-concepts. These two classes are "pos" and "neg", respectively meaning *positive* and *negative* qualities characterizing the intangible Non-concept. These classes should be linked directly to a socket, as seen in the examples below.

### 6.2   Pluggable Ontology: Intangible Non-Concept Examples

We here refer to intangible non-concepts, noun or adjective. In a typical example of natural language evolution seen in a news magazine (Economist, May 2014, [5]), an article referred to a "non-coup". It looked like a military coup, but the country army insisted that it is not a coup. The army took up positions in the capital city, but kept a light footprint. This is a non-concept with both a positive (coup) and a negative attribute (light footprint) relative to the original concept.

### 6.2.1 Example 1: Non-event

According to dictionary definitions, there are two possible meanings for a *Non*-event. The first definition is a "much publicized event, which is disappointing". The corresponding pluggable ontology is seen in Fig. 15.

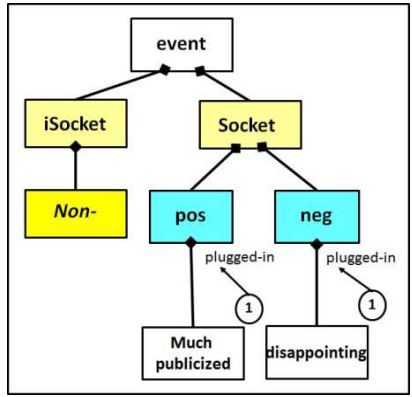

Figure 15: Non-event pluggable ontology – It is a non-event since positive and negative attributes have been plugged-in – both with cardinality value=1. The pos attribute is "much publicized". The neg attribute is "disappointing".

A second definition is a much publicized event, which does not occur. The corresponding pluggable ontology is almost the same as in Fig. 15, in which the "disappointing" attribute is substituted by "not occurring".

### 6.2.2 Example 2: Non-trivial

*Non*-trivial is an adjective with more than one dictionary definition. A mathematical one: "an expression in which at least one variable is not equal to zero". A second one states "having some importance", i.e. the subject has importance, but not much. The pluggable ontology of the second meaning is seen in Fig. 16.

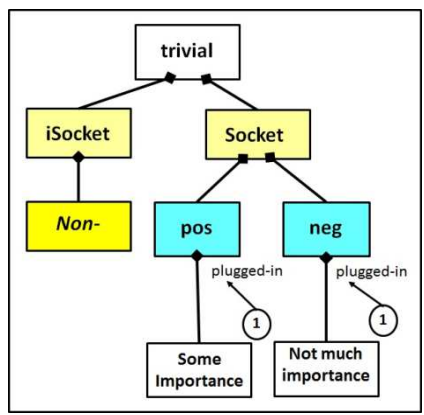

Figure 16: Non-trivial pluggable ontology – It is non-trivial since "some importance" (positive) and "not much importance" (negative) attributes have been plugged-in, both with cardinality value=1.

### 6.2.3  Example 3: Non-cooperation

The concept *Non*-cooperation has an interesting dictionary definition: "passive refusal to cooperate". The corresponding pluggable ontology is seen in Fig. 17.

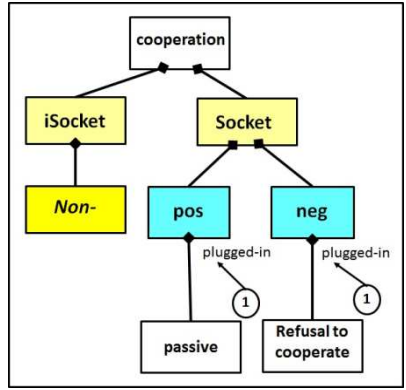

Figure 17: Non-cooperation pluggable ontology – It is non-cooperation since both "passive" (positive) and "refusal to cooperate" (negative) attributes have been plugged-in, both with value=1.

### 6.2.4  Example 4: Non-standard

We use the term *standard* as a noun to mean a "procedure or a product, that is widely recognized or employed". There is an adjective with a corresponding meaning, for instance, in the sentence "a standard textbook for a discipline".

Now, what is the meaning of the next sentence?

-"This is a non-standard procedure to attack the problem".

One is not saying that this is not a recognized procedure. One just means that it is not "widely" recognized. But such a sentence has a stronger meaning: there is an implicit dispute between a side refuting it as a suitable procedure and another side probably proposing it as a good candidate for a standard. Before the dispute is resolved, it has a non-standard status.

Within linguistics this dispute – of whether *non*-standard or some of its previous equivalents are adequate or unfairly judgmental – led to an alternative concept "sub-standard", which did not resolve the dispute.

The corresponding pluggable ontology is seen in Fig. 18.

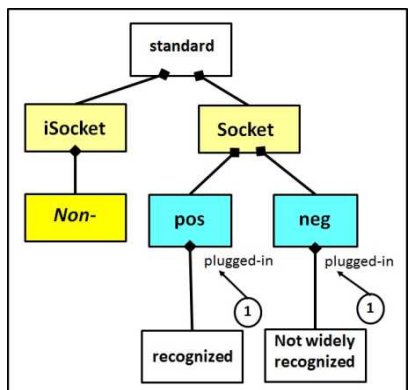

Figure 18: Non-standard pluggable ontology – It is non-standard since it is both "recognized" (positive) but "not widely recognized" (negative), with both these attributes plugged-in with value=1.

## 7   DSOCKET: A DYNAMICALLY MODIFIABLE DESIGN PATTERN

Summarizing, this work proposed a true Design Pattern for dynamically modifiable software systems in the highest abstraction design level. We call it the DSocket Design Pattern. DSocket is an abbreviation for Design Socket. The purpose of this Design Pattern is to enable representation by a concise pluggable ontology of a whole subsystem and its current modifications. The DSocket Pattern allows easy recognition of the modifiable subsystem and the nature of the modifications made. It is schematically shown in Fig. 19.

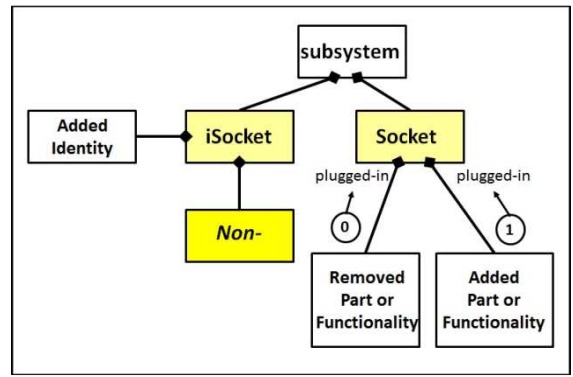

Figure 19: The DSocket Design Pattern – It refers to a known subsystem that can be modified by removal/addition of parts or functionalities in/out the Socket and addition of identities in/out the iSocket. *Non-* is a particular case of identity.

The DSocket Design Pattern has the following characteristics:
- *Whole modifiable subsystem represented by a single class* – whose class name is the *subsystem* name;
- *Abstract connectors* – it has two abstract connector classes, Socket and iSocket; it is strongly recommended to keep the names of these connectors to allow recognition of the DSocket Design Pattern in a larger system;
- *All the arrows stand for composition* – in contrast with the so-called GoF design patterns by Gamma et al. [18] in which the dominant relations between classes are inheritance, here the dominant relation is composition; composition is the recommended relation by the GoF book itself (page 20 in ref. [18]);
- *Unrestricted number of plugged-in entities* – one may add/remove any number of parts, functionalities or identities, although generally this number is a small integer.

## 8   DISCUSSION

### 8.1   Tangible and Intangible Non-Concepts Treated Equally

A central result of this paper is that *Non*-concepts for tangible and intangible things – subsystems – are treated equally by the same techniques. Intangible non-concepts appear naturally in the colloquial language that we speak.

The extension of the ideas to tangible *non*-concepts apparently is not so obvious, but it is plausible if one takes into account two basic arguments:
1. *Software Systems are expressed in Natural Language* – already in the Introduction of this paper we claimed that software systems in their highest abstraction levels are expressed in terms of natural language concepts, similarly to the intangible non-concepts.
2. *Software Systems are Ubiquitous* – nowadays software is widespread, and is a major component of embedded systems. Just to give a single example, software is a major component of cars travelling in our streets. The day may come that cars (fig. 7 in sub-section 4.2) will be an essentially software intangible subsystem covered by some metal carcass.

### 8.2  Advantages of Non-Concepts and their Pluggable Ontologies

We list three essential advantages of using Non-concepts and their ontology representation:

1) *Standard DSocket Modifiability Design Pattern* – we propose the systematic use of a standard design pattern for modifiable sub-systems; this has been intentionally repeated in the examples of this paper;
2) *Reduced Number of Standard Connectors* – the number of connector classes appearing in the standard pattern should be a minimal set; we have used just five such classes – *Non-, socket, isocket, pos, neg* – although their exact number is an open question (see below the Future Work sub-section 8.5);
3) *Preservation of Natural Language Meanings* – One should be familiar with Natural Language meanings and be able to express them in software systems' design. Our treatment of the meaning of *Non*-concepts in this paper is a contribution to this purpose.

### 8.3  Engineering Embedded and Software Systems

The pragmatic implications of *non-* concepts refer to design of systems of a few kinds:

a) *product lines* – with a *variety of models*, say a car or a printer that is marketed in various countries, may have different component parts in the final product, or different sources of parts manufacturing;
b) *removable parts* – for the sake of transportation, or avoiding worn parts before delivery;

In such cases, to enable system flexibility, one could use abstract sockets to explicitly manipulate parts with differing status, viz. to suitably label the respective parts along design, manufacturing and delivery stages, avoiding mistakes caused by lack of differentiation.

### 8.4  Between Identity and Composition

There have been disputing philosophical positions concerning whether identity is composition of parts or essentially different. Non- concepts imply that one cannot achieve a sharp resolution of this dispute.

The formulation of Non- concepts and sockets in this work and all the examples given lead us to the following position whether identity *is/isn't* composition. Identity and composition are entangled. To a certain extent, composition changes by parts' removal/addition do not affect identity. Beyond further removal/addition of parts, identity breaks down. This is not marked by a fixed quantitative limit; it depends on the part types and order of removal/addition. For instance, it is widely accepted from the conceptual point of view that a car turns into a non-car when the car engine is taken out of the car.

Art objects can be transformed into non-objects by other means, such as de-contextualizing. A well-known example is the concrete fountain put by Marcel Duchamp [33] in a museum. It created a scandal since its identity was immediately recognizable. The fountain was intact, but lost its intended functionality.

Art objects, such as Magritte's non-objects, trigger interesting discussions, allowing refinement of conceptualization issues. The question whether the non-pipe is a real object or just an image of an object, is not a real issue once one considers thought experiments like using Internet-Video which convert real objects into images.

Terminology issues regarding the most suitable denomination of non-objects: say quasi-objects or partial-objects, do not seem of basic importance. Once the ideas are accepted a suitable term will be found.

### 8.5  Future Work

The presentation of the work in this paper is largely informal. The most important future issue is a more formal treatment of *Non*-concepts. Our current preference is to base it on Formal Concept Analysis (FCA), perhaps augmented by Description Logics (DL).

A practical testing of Non-concepts and their pluggable ontologies may be facilitated by designing and implementing a plugin e.g. for the Protégé tool [27].

Finally some open questions regarding Non- concepts:

Do we need additional abstract connector classes to characterize *non-* concepts? While it is satisfactory that with a minimal set of generic classes – *Non-*, Sockets and iSockets, pos and neg – one has a quite flexible basis to build pluggable ontologies, one still needs a comprehensive investigation to provide a more definitive answer. In particular, intangible *non*-concepts deserve further investigation in this respect.

Are pluggable ontologies completely equivalent to modular ontologies? Sockets in the context of this paper, could be a natural mechanism to attach ontology modules.

## 8.6  Main Contribution

The main contribution of this work is the DSocket Design Pattern, the outcome of the analysis of <u>*Non*-concept</u> and its pluggable ontological representation. Non-concept is not a matter of degree. It is an entity of a new kind. A *non*-concept is neither a concept, nor a ¬concept.

## Acknowledgments

We are grateful for useful suggestions given by anonymous referees of our original paper (ref. [11]) presented in the KEOD'2012 Conference.

Non-Concept Software Subsystems: Tangible and Intangible	Iaakov ExmanConference, IC3K'2013, Vilamoura, Portugal, September 2013, pp. 260-275, Vol. 454 of Communications in Computer and Information Science, Springer, Berlin.
DOI = 10.1007/978-3-662-46549-3_17.

15. Exman, I. and Iskusnov, D., 2015. "Apogee: Application Ontology Generation with Size Optimization", in Fred, A., Dietz, J.L.G., Aveiro, D., Liu, K. and Filipe, J. (eds.) Knowledge Discovery, Knowledge Engineering and Knowledge Management, Revised Selected Papers of the 6[th] Int. Joint Conference, IC3K'2014, Rome, Italy, October 2014, pp. 477-492, Vol. 553 of Communications in Computer and Information Science, Springer, Berlin.
DOI = 10.1007/978-3-319-25840-9_29.

16. Ferre, S., 2006. "Negation, Opposition, and Possibility in Logical Concept Analysis", in Missaoui, R. and Schmidt J. (eds.) Formal Concept Analysis, 4[th] ICFCA Proc. Int. Conf. Dresden, Germany. LNCS Vol. 3874,Springer-Verlag, Berlin, Germany. DOI: 10.1007/11671404_9

17. Flouris, G., Huang, Z., Pan, J.Z., Plexousakis D. and Wache H., 2006. "Inconsistencies, Negations and Changes in Ontologies", Proc. AAAI'06 21[st] Nat. Conf. on Artificial Intelligence, Vol. 2, pp. 1295-1300, AAAI Press.

18. Gamma, E., Helm, R., Johnson, R. and Vlissides, J., 1995. *Design Patterns – Elements of Reusable Object-Oriented Software*, Addison-Wesley, Boston, MA, USA.

19. Guarino, N., 1998. "Formal Ontology and Information Systems", in Proc. of FOIS'98, Amsterdam, IOS Press, pp. 3-15.

20. Hois, J., Bhatt, M. and Kutz, O., 2009. "Modular Ontologies for Architectural Design", in Ferrario. R. and Oltramari, A. (eds.) Formal Ontologies Meet Industry, Vol. 198, of Frontiers in Artificial Intelligence and Applications, IOS Press. DOI: http://dx.doi.org/10.3233/978-1-60750-047-6-66

21. Hwang, C.H., 1999. "Incompletely and Imprecisely Speaking: Using Dynamic Ontologies for Representing and Retrieving Information", Proc. KRDB'99 6[th] Int. Workshop Knowledge Representation meets Databases, pp. 14-20.

22. Kripke, S., 1977. "*Identity and Necessity*", pp. 66-101, in Schwartz, S.P. (ed.) *Naming, Necessity and Natural Kinds*, Cornell University Press, Ithaca, NY, USA.

23. Lewis, D., 1993. "Many, But Almost One", in ref. (Bacon, 1993), pp.23-37.

24. Liao, Shen-yi, 2005. "Things are Their Parts", Logos, Vol. II, Issue 2, pp. 44-61 (Spring 2005).

25. Merricks, T., 1999. "Composition as Identity, Mereological Essentialism, and Counterpart Theory", Australasian Journal of Philosophy, 77, pp. 192-195.

26. Priss, U., 2006, "Formal Concept Analysis in Information Science", in Cronin, B, (ed.) Annual Review of information Science and Technology, Vol. 40, pp. 521-543. DOI: 10.1002/aris.1440400120.

27. Protégé - A free, open-source ontology editor and framework for building intelligent systems.
http://protege.stanford.edu/

28. Rector, A., Horridge, M., Iannone, L. and Drummond, N., 2008. "Use Cases for Building OWL Ontologies as Modules: Localizing, Ontology and Programming Interfaces & Extensions", in Proc. SWESE'08 4[th] Int. Workshop on Semantic Web Enabled Software Engineering.

29. Reiterer, E., Dreher H. and Gutl, C., 2010. "Automatic Concept Retrieval with Rubrico", in Schumann, M., Kolbe, L.M., Breitner, M.H. and Frerichs, A. (ed), Multikonferenz Wirtschaftsinformatik - MKWI 2010. Goettingen, Germany: Universitaetsverlag Goettingen.

30. Schlicht, A. and Stuckenshmidt, H., 2008, "Towards Distributed Reasoning for the Web", in Proc. WI-IAT'08 IEEE/WIC/ACM Int. Conf. on Web Intelligence and Intelligent Agent Technology, Vol. 01, pp. 536-539. DOI: 10.1109/WIIAT.2008.396

31. Simons, P., 1987. *Parts: a Study in Ontology*. Clarendon Press, Oxford, UK.

32. Varzi, A., 1996. "Parts, Wholes and Part-Whole Relations: The Prospects of Mereotopology", Data and Knowledge Engineering, Vol. 20, pp. 259. DOI: 10.1016/S0169-023X(96)00017-1

33. Wikipedia, 2015. Marcel Duchamp fountain, 1917.
https://en.wikipedia.org/wiki/Fountain_%28Duchamp%29
January 2016 / [Extended from KEOD Conf. paper in ref. [11] / October 2012]	Page 17